\documentclass[12pt]{article}
\newcommand{\rof}[1]{(\ref{eq:#1})}
\date{}
\title{Newton's Second Law in a Noncommutative Space}
\author{Juan M. Romero, J.A. Santiago and J. David Vergara \\
\\
Instituto de Ciencias Nucleares, U.N.A.M.,\\
 Apdo. Postal 70-543, M\'exico D.F., M\'exico\\
\\
sanpedro, santiago, vergara@nuclecu.unam.mx }
\begin{document}
\maketitle

\abstract{ In this work we show that corrections to the Newton´s
second law appears if we assume that the phase space has a
symplectic structure consistent with the rules of commutation of
the noncommutative quantum mechanics. In the central field case we
find that the correction term breaks the rotational symmetry. In
particular, for the  Kepler problem, this term takes the form of a
Coriolis force produced by the weak gravitational field far from a
rotating
massive object.}




\section{Introduction}

One of the subjects in which physicist have taken more
attention in the last years is in the study of the
noncommutative spaces. These are characterized such that their
coordinate operators satisfy the relation
\begin{equation}
[\hat x^{i},\hat x^{j}]=i\hbar \Theta^{ij}.
\end{equation}
Here the constant parameter of the noncommutativity is given by
$\hbar \Theta^{ij}$ that is real, antisymmetric and has units of
area. There are several reasons for which physicist are interested
in these spaces. For example, the quantum Hall effect, that is one
of most studied systems in condensed matter, presents
noncommutativity in the canonical coordinates and momenta
\cite{eza:gnus}. On the other hand, in string theory, under
certain background, we have the noncommutativity in the edges of
the open strings, and therefore in the coordinates of $D$-branes
\cite{W:gnus}. In addition a new field theory can be constructed
if one changes the standard product of the fields by the star
product (Weyl-Moyal):
\begin{equation}
(f*g)(x)=\exp
(\frac{i}{2}\Theta^{ij}\partial_{i}\partial_{j})f(x)g(y)|_{x=y},
\end{equation}
where $f$ and $g$ are infinitely differentiable functions.
In this theory some interesting results have been found
\cite{Sz:gnus}, for example, it was shown that there is a relation between
the infrared and ultraviolet divergences \cite{min:gnus}. Practically all
the interactions have been put in this language,
except the gravitational one. In addition, assuming the commutation rules:

\begin{equation}
[\hat x^{i}, \hat x^{j}]=i\hbar \Theta^{ij}, \quad [\hat x^{i}, \hat p_{j}]=i \hbar
\delta ^{i} _{j}, \quad [\hat p_{i}, \hat p_{j}]=0.\label{eq:>}
\end{equation}
a noncommutative quantum mechanics can be constructed, of which some
relevant results have been obtained \cite{chi:gnus}\cite{Ners:gnus}.\ \

In this work we will assume that we have a symplectic structure
consistent with the commutation rules \rof{>} and we will obtain
the corresponding equations of motion. We show that there is a
correction to the Newton's second law. We will see that this
correction turns out to be proportional to the noncommutative
parameter and also to the potential of the model. Thus, this new
force can be seen as the result of a perturbation in the space
caused by an external field. We also show that in the case of a
central field potential the correction can be interpreted like the
analog of a Coriolis force. One of the well known characteristics
of the noncommutative systems is that the Lorentz symmetry  is
broken \cite{Ar:gnus}, in our case the correction to the Newton's
second law  breaks the symmetry under rotations\footnote{A different point of view about the breaking of the rotational invariance 
can be found in \cite{Deri:gnus}.}. We will see two
concrete examples, firstly, the potential of a $3$-dimensional
harmonic oscillator and secondly the Kepler problem. For the
harmonic oscillator we obtain equations of motion that can be seen
as those of an oscillator in a background constant magnetic field
of the order of the noncommutative parameter. For the Kepler
problem, the term we have  obtained has the form of a Coriolis
force as produced by a very far gravitational field of a rotating
massive object. \ \

\section{Noncommutative Classical \\
 Mechanics }

To begin with, we suppose that we have a set of variables
${\zeta^{a}}$, with $a=1, ...,2n$, and an antisymmetric matrix
$\Lambda^{ab}=\{\zeta^{a},\zeta^{b}\}$. Given  $F$ and $G$
functions of ${\zeta^{a}}$, we can define a symplectic structure
as \cite{Mars:gnus}

\begin{eqnarray}
\{F,G\}=\{\zeta^{a},\zeta^{b}\}\frac{\partial F}{\partial\zeta^{a}}\frac{\partial G}
{\partial\zeta^{b}}.
\end{eqnarray}
In terms of this structure and given a Hamiltonian $H=H(\zeta^{a})$
we can write the equations of motion as
\begin{eqnarray}
\dot \zeta^{a}=\{\zeta^{a}, H \}.
\end{eqnarray}
More in general, for any function $F$ defined in this space we have
\begin{eqnarray}
\dot F=\{F, H \}.
\end{eqnarray}
On the follow, we will consider the phase space given by
$\zeta^{a}=(x^{i}, p_{i})$, with $i=1,2,3$. \ \ Let us now
consider the rules of commutation of noncommutative quantum
mechanics \rof{>}, the symplectic structure consistent with these
is defined by:
\begin{equation}
\{ x^{i }, x^{j}\}=\Theta^{ij }, \quad \{x^{i }, p_{j}\}= \delta
^i_ {j}, \quad \{p_{i }, p_{j}\}=0. \label{eq:. }
\end{equation}
As mentioned  previously, $\hbar\Theta^{ij}$ must have dimensions
of area, and therefore by assuming that this parameter is of the
Planck's area order, $l^{2}_{p}=\frac{\hbar G}{c^{3}}$ the tensor
$\Theta^{ij}$ must be of $\frac{G}{c^{3}}$ order. Thus, in the
classical limit, the symplectic structure will not have $\hbar $,
as should be. \ \ On the other hand, taking  $F$ and $G$, both two
arbitrary functions defined on the phase space and using \rof{. }
we can obtain the following modified Poisson brackets
\begin{eqnarray}
\{F,G\}=\Theta^{ij}\frac{\partial F}{\partial x^{i}}\frac{\partial
G}{\partial x^{j}}+ (\frac{\partial F}{\partial
x^{i}}\frac{\partial G}{\partial p_{i}}-\frac{\partial F}
{\partial p_{i}}\frac{\partial G}{\partial x^{i}}).
\end{eqnarray}
We now consider a Hamiltonian with the form
$$H=\frac{p_{i}p^{i}}{2m}+V(x), $$
the equations of motion corresponding to this symplectic structure
are given by
\begin{eqnarray}
\dot x^{i}=\frac{p^{i}}{m}+\Theta^{ij} \frac{\partial V}{\partial
x^{j } }
\label{eq:,}\\
\quad \dot p_{i}=-\frac{\partial V}{\partial x^{i}},
\label{eq:h1 }
\end{eqnarray}
which can be written as
\begin{equation}
m\ddot x^{i}=-\frac{\partial V}{\partial x_{i}}+
m\Theta^{ij}\frac{\partial^{2 } V } { \partial x^{j}\partial
x_{k}} \dot x_{k }. \label{eq:.. }
\end{equation}
This equations are to be new Newton's second law.
In the second term of \rof{.. }
we can see a correction due to the noncommutative rule.
This new term is generated by both, the background space,
through the factor of the noncommutativity, an also for variations
in the potential.
The external field
produces a perturbation in the space that induces
this new force. The equations \rof{.. } has been obtained in two 
dimensions in \cite{Acatri:gnus}.

\section{Central Field}

In the case of a central potential $V(x)=V(r)$ the correction to
the second Newton's law can be written in a more suggestive form.
Let us consider the tensor $\Theta^{ij}=\epsilon^{ijk }
\Theta_{k}$, that has been used to study noncommutative systems at
level of quantum mechanics \cite{chi:gnus}. Then, for the central
potential case the momenta, defined by the equation \rof{,} has
the form
\begin{equation}
p^{i}=m\dot x^{i}+m\epsilon^{ijk}\Omega_{j}x_{k} \quad { \rm
with}\quad \Omega_{j}=\frac{1}{r}\frac{\partial V}{\partial
r}\Theta_{j },
\end{equation}
this equation represent the momenta of a  particle as seen from a
non-inertial system with angular velocity $\Omega_{j}$
\cite{lan1:gnus}. The equation \rof{.. } for the case that we are
considering is given by:
\begin{equation}
m\ddot x^{i} =-\frac{x^{i}}{r}\frac{\partial V}{\partial r}+
m\epsilon^{ijk}\dot x_{j}\Omega_{k}+ m\epsilon^{ijk}x_{j}\dot
\Omega_{k}.\label{eq:...}
\end{equation}
the second term in the right hand side of this equation is the
analog of an inertial force produced by the non-uniformity of the
rotation, whereas the third one it is the Coriolis force caused by
the same rotation. The correction terms clearly break the
rotational invariance under of the central field.

Let us now consider two examples of central field. First we
consider the potential of a three-dimensional harmonic oscillator
$V(r)=\frac{\omega^{2}}{2}r^{2}$. In order to simplify the
calculations take $\Theta_{i}=\delta_{i3}\Theta$. Then
\begin{equation}
m\ddot x^{i}=-\omega^{2}x^{i}+ m\omega^{2}\Theta
\epsilon^{ij3}\dot x_{j}.\label{eq:os }
\end{equation}
By the Larmor theorem we can see how the perturbation in the space, caused by
the external potential, produces a kind of constant
magnetic field
in the direction of the vector $\Theta_{i}$. Equation \rof{os } has well-known
solution: on the axis $x_{3}$ we have oscillations of frequency $\omega$, whereas
in the perpendicular plane to this axis the frequency of the
oscillations to first order in $\Theta$ has the form $\omega_{\Theta}\approx \omega
(1\pm \frac{\Theta m \omega}{2}). $
This situation is similar to the quantum case \cite{poly:gnus} .\\

In the case of the Kepler potential
$V(r)=\frac{\alpha}{r}$,
the angular velocity takes the form
$$\Omega_{i}=-\frac{\alpha}{r^{3}}\Theta_{i}, $$
and the equations of motion are given by
\begin{equation}\label{cfK}
m\ddot x^{i}=\frac{x^{i}}{r}\frac{\alpha }{r^{2 } } - \frac{\alpha
m}{r^{3 } } \epsilon^{ijk}\dot x_{j}\Theta_{k}+
m\epsilon^{ijk}x_{j}\dot \Omega_{k}.
\end{equation}

By noting the $r$ dependence of the angular velocity we see that
the resulting Coriolis force is analog to that produced by a
gravitational field of a massive rotating object
\cite{lan2:gnus,ciu:gnus}, where the $\Theta_{i}$ vector plays a
similar  role to the angular momenta of the object, whereas the
last correction can be seen as a force
produced by a non-uniform rotation of it \cite{io:gnus}.
A perturbative analysis of (\ref{cfK}) has been obtained very recently 
in \cite{Mirza:gnus}. \\

By using the equation of motion \rof{...}, we can see that the
angular momenta $L^{i}=m\epsilon^{ijk}x_{j}\dot x_{k}$ is not
conserved. In addition, from \rof{. } is possible to show that it
does not generate rotations. Nevertheless we can construct the
amount
$$L_{\Theta}=\Theta^{ij}x_{i}p_{j}+\frac{1}{2 } \Theta^{ij}p_{j}\Theta_{ik}p^{k}, $$
this one generates the transformations:
\begin{equation}
\delta x^{i}=\{x^{i}, L_{\Theta}\}= \Theta^{ik}x_{k}, \quad \delta
p_{i}=\{p_{i}, L_{\Theta}\}= -\Theta_{ik}p^{k}.\label{eq:o}
\end{equation}
In the central field case $L_{\Theta}$ is conserved.
Also, we can generate rotations in the $\Theta_{i}$ directions by replacing
$\Theta^{ij}=\epsilon^{ijk}\Theta_{k}$ into the transformations \rof{o}.\\

We note that the equation \rof{...} is also correct in the case that the
noncommutative parameter depends on the coordinates. For example, there is no corrections to the equations of motion in the classical Fuzzy sphere case, where we have
\begin{equation}
\{x^{i},x^{j}\}=\Theta \epsilon^{ijk}x_{k},
\end{equation}
beyond this particular case, we are currently doing progress in this direction.

\section{Conclusions }
In this work we find the corrections to the second Newton's law
for consider the symplectic structure consistent with the
commutation rules of noncommutative quantum mechanics. It is
interesting to observe that the correction term is proportional to
both, the field and the noncommutative parameter. This allows us
to interpret the resulting force as an effect caused by the
perturbation of the external field on the space. Another
interesting result we have obtained is for the case of a
$3$-dimensional oscillator. The equations of motion can be seen as
those of an oscillator in a background constant magnetic field.
For the Kepler's problem we have obtained a force as produced by a
weak gravitational field far from a rotating object. In addition,
we have found that in a central field the correction term breaks
the rotational symmetry, nevertheless the generator of rotations
is conserved
in the direction of the noncommutative parameter. 
\\

\section{Acknowledgments}
The authors acknowledge partial support from  DGAPA-UNAM grant
IN117000 (J.M.R. and J.D.V.).

\end{document}